# Effect of the troposphere on surface neutron counter measurements

K.L. Aplin (a), R.G. Harrison (b) and A.J. Bennett (b)

(a)     Space Science and Technology Department, Rutherford Appleton Laboratory, Chilton, Didcot, Oxon, OX11 0QX UK (k.l.aplin@rl.ac.uk)
(b)     Department of Meteorology, The University of Reading, PO Box 243, Earley Gate, Reading, Berks, RG6 6BB UK (r.g.harrison@rdg.ac.uk, a.j.bennett@rdg.ac.uk)

**Abstract**

Surface neutron counter data are often used as a proxy for atmospheric ionisation from cosmic rays in studies of extraterrestrial effects on climate. Neutron counter instrumentation was developed in the 1950s and relationships between neutron counts, ionisation and meteorological conditions were investigated thoroughly using the techniques available at the time; the analysis can now be extended using modern data. Whilst surface neutron counts are shown to be a good proxy for ionisation rate, the usual meteorological correction applied to surface neutron measurements, using surface atmospheric pressure, does not completely compensate for tropospheric effects on neutron data. Residual correlations remain between neutron counts, atmospheric pressure and geopotential height, obtained from meteorological reanalysis data. These correlations may be caused by variations in the height and temperature of the atmospheric layer at ~100hPa. This is where the primary cosmic rays interact with atmospheric air, producing a cascade of secondary ionising particles.

**1. Introduction**

The possible effect of cosmic rays on Earth's climate system is controversial, as many physical mechanisms have been proposed, but none proven. In the last few years, evidence for two principal mechanisms by which atmospheric ion formation by cosmic rays could affect Earth's climate has emerged. These are described in e.g. Harrison and Carslaw (2003) and are briefly summarised below:
1. ion-mediated particle formation, which may lead to the growth of cloud condensation nucleii (e.g, Svensmark and Friis-Christensen, 1997; Yu and Turco, 2001)
2. enhanced removal of charged droplets from clouds (e.g., Tinsley, 2000; Tripathi and Harrison, 2002)

Theoretical work, model predictions and laboratory experiments exist supporting these hypotheses, but results from relevant atmospheric experiments are still relatively sparse. Accumulation of further atmospheric data to corroborate or refute the existence of any of these effects requires quantification of ion concentrations in Earth's troposphere and stratosphere. Ideally, direct investigation of ionisation effects on climate would require frequent vertical profiles of meteorological and ion measurements over a wide spatial area. Whilst new atmospheric ion instrumentation is becoming available (Aplin and Harrison, 2001; Holden, 2003), ion data are still relatively rare. Routine ion measurements at the surface exist at very few locations (e.g. Hõrrak *et al*, 2000), and ion data outside the atmospheric boundary layer are even sparser. This is the reason why the cosmic ray flux, responsible for almost all atmospheric ionisation except close to the continental surface, is often used as a proxy for atmospheric ion concentrations. Cosmic ray fluxes have been routinely monitored by a surface neutron counter network since the International Geophysical Year in 1956/7. There are now ~50 monitors well-distributed worldwide, and the data is readily available on the Internet (Pyle, 2000).

Many studies of cosmic ray effects on meteorological parameters have been based on data from this neutron monitor network. These studies rely on the assumptions that neutron monitor data are, first, a good proxy for the ionisation rate from cosmic rays and, secondly, independent of atmospheric parameters. Whilst the validity of these assumptions was established for the purposes needed at the time during the development of the neutron counter fifty years ago, investigations have not been extended using modern techniques. Many of the possible cosmic ray-climate signals appear to be at the few percent level (e.g. Svensmark and Friis-Christensen, 1997), so there is a need to verify that such signals are not spuriously generated by residual atmospheric effects on the cosmic ray measurements, incompletely removed by the atmospheric corrections. In this paper, historical and modern data are used to develop the early work relating atmospheric processes to surface cosmic ray intensities. Background into cosmic ray propagation in the atmosphere is given in Section 2, followed by a description of the geophysical and meteorological parameters that were understood in the 1950s to modulate cosmic radiation. In Section 4 the development of cosmic ray neutron monitoring and the routine correction for atmospheric pressure are briefly discussed. Finally, modern meteorological data is used to show that there are residual tropospheric effects on pressure corrected neutron counts, which may cause correlations between cosmic ray fluxes and other atmospheric parameters.

2. **Cosmic rays in the atmosphere**

Most atmospheric ions originate from cosmic ray ionisation. Near the continental surface, natural radioisotopes contribute about 80% of the ionisation rate, defined as the number of ions formed per unit volume per second, but this contribution decreases with height and is negligible outside the atmospheric boundary layer. Cosmic rays are high-energy ionising radiation, entering Earth's atmosphere from space; most are now thought to come from supernovae (Shaviv, 2002; Wolfendale, 2003). Primary cosmic rays are energetic particles, which interact with atmospheric air molecules when the air becomes sufficiently dense, at a pressure surface of ~100-200 hPa (~100-200 gcm$^{-2}$). These pressures are found at ~11-16 km in the upper troposphere or lower stratosphere, depending on season and latitude. When cosmic rays interact with other air molecules, secondary subatomic particles, usually mesons, are produced, which interact with more air molecules, and so on as the air density (pressure) increases down to the surface, to produce a "nucleonic cascade" of particles including high-energy and thermal neutrons, protons, muons and electrons (Simpson *et al.*, 1953). Thermal neutrons interact strongly with atmospheric water vapour, whereas the protons, muons and electrons lose energy by ionisation when they interact with atmospheric air molecules. The ionisation rate therefore decreases from a maximum at ~15 km, where the flux of secondary particles first exceeds the primary particle flux, to the top of the boundary layer where ionisation from natural radioactivity starts to dominate over land. Cosmic rays remain the dominant source of ionisation in the oceanic boundary layer.

**3. Modulation of atmospheric ionisation**

3.1 Geophysical effects

Simpson (2000) identified the selection and modulation of cosmic ray energies by geomagnetic latitude from measurements in the 1950s. The cosmic ray energy spectrum extends over ~1-$10^{21}$eV, with the particle flux dropping off as energy increases (Bazilevskaya, 2000). The lower-energy cosmic rays are selectively screened by Earth's magnetic field at the mid-latitudes and near the equator. Above geomagnetic latitudes of ~50°, the cosmic ray screening is insensitive to latitude (Bazilevskaya, 2000). Solar activity affects cosmic rays, as the magnetic field irregularities in the solar wind deflect cosmic rays away from Earth, so that primary cosmic ray penetration into the atmosphere is highest at solar minimum. Tropospheric ionisation rates are greatest at high latitudes during solar minimum, and lowest at equatorial latitudes at solar maximum (Gringel *et al.*, 1986).

3.2 Meteorological effects (1940s view)

Meteorological effects on cosmic rays in the atmosphere were discovered during fundamental studies of atmospheric structure and properties in the first half of the twentieth century. For example, Loughridge and Gast (1939, 1940) observed weather fronts affecting surface ionisation chamber measurements (see Section 4) on a cruise in the North Pacific. Their expedition was intended to investigate latitudinal variations of cosmic ray intensity between Seattle and Alaska, but a relationship between cosmic ray ionisation and the passage of fronts was also detected. Cold fronts cause a 1% decrease, and warm fronts a 0.5% increase in ionisation over the 30 hours it took for the fronts to pass. This effect was robust, even after the data had been corrected for surface air pressure changes. Blackett (1938) had predicted the existence of a meteorological effect on cosmic rays due to variations in the average height, and therefore temperature, of the atmospheric layer where the primary particles interact to produce secondary ionising particles. This was referred to as the "mesotron producing layer" in Loughridge and Gast (1939), before the term meson was universally used. Changes in temperature affect meson range in air, and influence the propagation of the nucleonic cascade, which modulates the tropospheric ionisation rate. Before the advent of widespread regular meteorological soundings, weather data above the surface were rare. Loughridge and Gast used assumptions based on early sounding data to infer pressure and temperature aloft during their experiment. Changes in the ionisation rate based on the thickness of the "mesotron producing layer" were estimated from the changing height of the tropopause, affected by weather fronts. These predictions fitted observed ionisation rate variations. Although ionisation chambers for cosmic ray monitoring are now obsolete, meteorological effects on modern cosmic ray data remain. Modern cosmic ray instrumentation, and meteorological effects upon contemporary cosmic ray data, will now be outlined.

**4. Surface cosmic ray detection instrumentation**

Cosmic rays were initially detected using ionisation chambers: sealed containers containing gas at atmospheric pressure, with a collecting electrode biased to a fixed potential, and connected to an electrometer. Radiation passing through the chamber creates ion pairs, one polarity of which is attracted to the collecting electrode. The current detected is proportional to the number of ion pairs created (e.g. Smith, 1966). This was the first technique to measure radioactivity, and early measurement units such as the Roentgen were related to the number of ions formed in air. Hess, who is credited with discovering cosmic rays, defined a unit for cosmic ray intensity, $I$, the volumetric cosmic ray ion production rate in nitrogen at standard temperature and pressure (Hess, 1939).

An example of the use of ionisation chambers was on the geophysical and atmospheric electrical research ship, the *Carnegie*. The ionisation chamber measured "penetrating radiation", i.e the surface ionisation rate from cosmic rays, by counting the number of ions produced per unit volume per second. The ionisation chamber used on the *Carnegie* was a copper chamber of about 22 litres in volume, larger than those commonly used at the time (Ault and Mauchly, 1926). Penetrating radiation measurements on cruises IV and VI, between 1915 and 1921, have been digitised here, and the geographic coordinates recorded in the measurement log converted into geomagnetic coordinates. This calculation required the location of the geomagnetic North Pole. The location of this pole, which hardly varied between 1915-1921, was calculated using altitude adjusted corrected geomagnetic coordinates (see http://www.wdc.rl.ac.uk/cgi-bin/wdcc1/coordcnv.pl). Coordinate conversion was achieved by modelling the Earth's magnetic field as a dipole in a spherical shell (Ziegler, 1996). The variation of the ionisation rate from "penetrating radiation" with geomagnetic latitude is shown in Figure 1. The ionisation rate increases with geomagnetic latitude, as expected from the Earth's magnetic field screening out lower energy cosmic rays towards the geomagnetic equator. Cruise VI made measurements over a wider geomagnetic latitude range than Cruise IV, and a flattening of the trace can be seen at geomagnetic latitudes >~50º, where the cosmic ray atmospheric penetration is no longer geomagnetically screened (see Section 3.1). As Figure 1 demonstrates, the ionisation chamber was a simple and effective instrument for measuring the ions formed by cosmic rays. Its disadvantages were that it excluded some of the lower energy particles (Simpson, 2000), and could also be subject to contamination from radioactivity in the walls of the ionisation chamber.

Simpson invented the neutron counter, which responds to the fast neutrons produced by the atmospheric interactions of cosmic rays (Simpson, 2000). In summary, the neutron counter is a boron trifluoride proportional counter with an enriched $^{10}B$ component. $^{4}He^{2+}$ is produced when a neutron interacts with $^{10}B$, and the doubly charged helium atoms are detected in the proportional counter. In an early comparison, ionisation chambers and the new neutron counters were flown together on an aircraft to calibrate the neutron counter (Simpson, 2000; Biehl *et al.*, 1949). Figure 2a shows ionisation rate and neutron counts at 9km, over a range of geomagnetic latitudes (Simpson, 2000). This indicates that there is an approximately linear relationship between ionisation rate and neutron counts in the troposphere over a geomagnetic latitude range of 10-50º. The relationship is not expected to differ at geomagnetic latitudes >~50º, for the reasons described in Section 3.1. The close relationship between ionisation chamber and neutron counter data is corroborated by results from the overlapping period 1953-1957 when an ionisation chamber was run at Cheltenham, Massachusetts (Ahluwalia, 1997) at the same time as the neutron monitor at Climax, Colorado, Figure 2b. Figure 2c shows a time series for the overlapping period and up to 2000, in which the solar cycle is clearly visible. As suggested in Section 3.2, a pressure correction was needed to compensate for the changes in ambient air pressure affecting cosmic ray propagation. With this pressure correction applied routinely, the neutron counter has become the standard surface instrumentation for cosmic ray monitoring.

4.1 Neutron counter pressure correction

The properties of the nucleonic cascade in the atmosphere depend on the interaction cross-section of atomic nucleii in air per unit volume. Surface cosmic ray intensity detected by any instrument is therefore affected by the integrated air density, or air pressure, in the column of air above it. The level at which primary particles interact with air is an especially important factor. The number of subatomic particles produced by the nucleonic cascade is a function of the distance travelled by the secondary mesons, and the meson range and lifetime is related to the temperature and pressure of the atmospheric layer where the primary particles interact (Sandström, 1965).

Early pressure corrections were based on linear regression using surface pressure measurement, the coefficients of which differed slightly for each station due to geomagnetic latitude variations. Use of the station surface pressure for the correction initially appears to have been a pragmatic choice based on the data available at the time. Sandström (1965) argued that the pressure correction could be improved by including data for the pressure and temperature of many atmospheric layers. This would take account of subtle variations in atmospheric structure affecting subatomic particle interactions (Sandström, 1965; Clem and Dorman, 2000). More recently, sophisticated techniques have been developed to compute the pressure correction from full Monte Carlo simulations of the nucleonic cascade (Clem and Dorman, 2000). Despite this advance, the neutron monitor stations retain a simple linear correction using surface pressure readings. Figure 3 shows the effect of the pressure correction at the Oulu neutron monitor (65.05ºN, 25.47ºE) for a sample year of neutron data. It is evident that the correction removes much of the pressure dependence of the neutron counter data, but there is still a small residual effect (for 2001) of ~ –2 counts/min/hPa or ~ -0.3% neutrons/% pressure change, with $r$=0.13. This may not be significant for some analyses, but, in the recent studies of cosmic rays and climate, the effects observed in data are often at the few percent level. In this context it is important to eliminate any residual effects of the atmosphere on the data used to represent cosmic ray intensity.

## 5. Atmospheric modulation of modern surface neutron count data

Meteorologists use numerical models of the atmosphere for data assimilation: this is the generation of globally gridded data coverage of past atmospheric properties derived as the best estimate of the atmospheric state from all observations. This reanalysis data gives atmospheric parameters available in height profiles from the surface to 10hPa, in 2.5º grid squares. The NCEP/NCAR reanalysis geopotential height (height of a pressure surface) data have been used to investigate meteorological effects on surface neutron counts. As the primary particles interact at a constant atmospheric pressure, the 100hPa geopotential height can be used as an indicator of the height of the meson producing layer (Sandström, 1965).

More information about the spatial variation of the relationship between geopotential height and surface neutron counts can be obtained from vertical meridional cross-sections. Neutron data from a midlatitude station, Climax (39.37ºN, 253.82ºE) were chosen, as it could be approximately centred on a latitudinal cross-section from 10ºS to 60ºN, covering the geomagnetic equator to the latitudes where the cosmic ray energies are no longer a function of latitude. Correlations between geopotential height and surface neutrons were computed from the surface (1000hPa) to the meson producing layer (100hPa). The correlation contours for 47 years of monthly average data are shown in Figure 4, indicating an anticorrelation at ~100hPa, at its highest over the tropics, with a maximum of $r = -0.43$.

The anticorrelation is expected for the following reasons. If the 100hPa geopotential height increases, the meson producing layer is likely to be colder, and the mesons have a shorter lifetime, which results in fewer interactions with air and a lower surface neutron count (Sandstrom, 1965). Whilst the ratio of meson interactions to meson decays is dominant (Clem and Dorman, 2000), there are several competing mechanisms related to the lifetime of unstable species in the atmosphere, and the lower troposphere temperature can also affect meson losses due to ionisation (Olbert, 1953).

The region of highest anticorrelation is close to the geomagnetic equator, where only the highest energy primary cosmic rays can enter the atmosphere. This suggests that the neutron production from high-energy nuclear disintegrations may be more sensitive to atmospheric effects than lower energy interactions. The meson producing layer is usually in the troposphere near the equator, and in the stratosphere in the mid to high latitudes, so the spatial variation of the anticorrelation may also be related to dynamical processes in the atmosphere.

This could be linked to the persistent relationship between the 10.7cm solar flux, a solar activity indicator, and the 30hPa geopotential height (Labitzke, 2001). However, it is difficult to separate the effects of cosmic ray and solar flux variations using purely statistical approaches. As mentioned in Section 3.1, cosmic ray intensity in the atmosphere is in antiphase with solar activity, but the 10.7cm solar flux is an indicator of total solar irradiance, and is greater at solar maximum. Cosmic ray intensity at Earth and solar activity indicators are therefore closely inversely correlated. This is illustrated in Figure 5, in which, following van Loon and Labitzke (2000), correlations with the 30hPa geopotential height are compared for 1958-1998. Figure 5a) reproduces the results in Figure 3 of van Loon and Labitzke (2000), for the spatial correlation variation between the 10.7cm solar flux and the 30hPa height. In Figure 5b), the pressure corrected monthly surface neutron counts from the University of Chicago's monitor at Climax are used instead. Figure 5a) and Figure 5b) are, as expected, inversely correlated, although there are small differences in position of the regions of high correlation. The magnitude of the correlation between neutron counts and geopotential height is also slightly greater. In each case there are physical reasons for expecting a correlation: the UV component of the solar irradiance is thought to modulate stratospheric dynamic processes through ozone changes (Haigh, 2003). Detailed theoretical predictions for both postulated mechanisms are needed to distinguish between the solar flux and cosmic ray effects.

## 6. Conclusions

Two assumptions about cosmic rays are frequently made in considering relationships between atmospheric ionisation and meteorological processes. The first assumption is that surface neutron counts are a good proxy for the atmospheric ionisation rate. Analysis of historical data comparing the neutron counter and ionisation chamber responses on a plane flying over geomagnetic latitudes of 10-50º indicated an approximately linear relationship between them. This readily confirms that surface neutron counts can be used as a proxy for ionisation rate. The 1950s data are useful because this is the only period where there is an overlap between regular cosmic ray measurements by both neutron counters and ionisation chambers. It could be valuable for contemporary studies of cosmic ray effects in the atmosphere to be able to calibrate surface neutron counts to the integrated ionisation rate. This would be possible by comparing, for example, balloon ascents measuring the ionisation rate (Bazilevskaya, 2000) with the colocated surface neutron count rate.

The second assumption is that the modulation of cosmic rays detected at the surface by atmospheric properties can be ignored. Physical mechanisms of atmospheric effects on cosmic ray propagation, modulated by meteorological factors, were established by the 1940s, but are perhaps not well known to contemporary climate scientists. It has been shown in Sections 4 and 5, using both simple surface measurements, and reanalysis geopotential height data, that the pressure correction used at the Oulu neutron monitors does not completely remove meteorological effects on surface neutron counts. Some of the variance in pressure corrected neutron data can be attributed to geopotential height variations affecting the properties of the meson producing layer. The residual effect is small, <1%, but may be highly spatially variable; it is likely to be greater at some locations. This is supported by spatial correlations between the 30hPa and 100hPa geopotential heights, and cosmic rays measured at the Climax neutron monitor station. Any mechanisms linking cosmic rays and climate may be similarly subtle and variable in magnitude; it is therefore necessary to understand what fraction of the variability remains from the effect of the atmosphere on cosmic rays, before quantifying the effects of cosmic rays in the atmosphere.

Globally gridded meteorological reanalysis data are now routinely and freely available, and would benefit from broader use within the geophysical sciences. It would be relatively straightforward to retrieve pressure and temperature profiles for specific neutron monitor stations. These could be used in conjunction with Monte Carlo simulations of the nucleonic cascade to correct for atmospheric effects on neutron counts more accurately. Without this further theoretical work, a small possibility exists that correlations found between solar and atmospheric parameters could include a component of the effects of the atmosphere on cosmic rays. However, the physical arguments based on ion-aerosol mechanisms appear generally persuasive.

**Acknowledgements**

KLA acknowledges partial funding from the UK Natural Environment Research Council (NERC New Investigators' Award NER/M/S/2003/00062), and the UK Particle Physics and Astronomy Research Council (PPA/G/O/2003/00025). AJB acknowledges a NERC studentship. RGH acknowledges a Visiting Fellowship at Mansfield College, Oxford. NCEP Reanalysis data were provided by the NOAA-CIRES Climate Diagnostics Center, Boulder, Colorado, USA, from their Web site at http://www.cdc.noaa.gov/. Cosmic ray data were obtained from the Oulu (http://cosmicrays.oulu.fi/) and Climax (http://ulysses.sr.unh.edu/NeutronMonitor/neutron_mon.html) neutron monitor stations. The Climax station is funded by National Science Foundation Grant ATM-9912341.

**Figure Captions**

Figure 1. Variation of the ion production rate due to "penetrating radiation" with geomagnetic latitude in the Southern Hemisphere. Measurements were made on the Carnegie cruises IV (1915) and VI (1921).

Figure 2. Neutron counter-ionisation chamber comparisons a) Calibration of one of the first neutron counters at 9km, over 10-50º geomagnetic latitude, after Simpson (2000) and Biehl *et al.* (1949) b) Calibration of neutron monitor to ionisation chamber (IC) for 1953-1957 c) Time series of Cheltenham (USA) ionisation chamber (Ahluwalia, 1997) and Climax neutron monitor.

Figure 3. Comparison of pressure corrected and uncorrected neutron count data at Oulu for 2001, as a function of surface atmospheric pressure.

Figure 4. Correlations between monthly averaged Climax corrected neutron data and geopotential height over –10ºS to 60ºN and for 1000-10hPa, for longitudes 252.5-255ºE.

Figure 5. Spatial correlations between 30hPa geopotential height, 1958-1998, and a) the 10.7cm solar flux, as in van Loon and Labitzke (2000), b) with the monthly surface neutron counts from the Climax neutron counter.

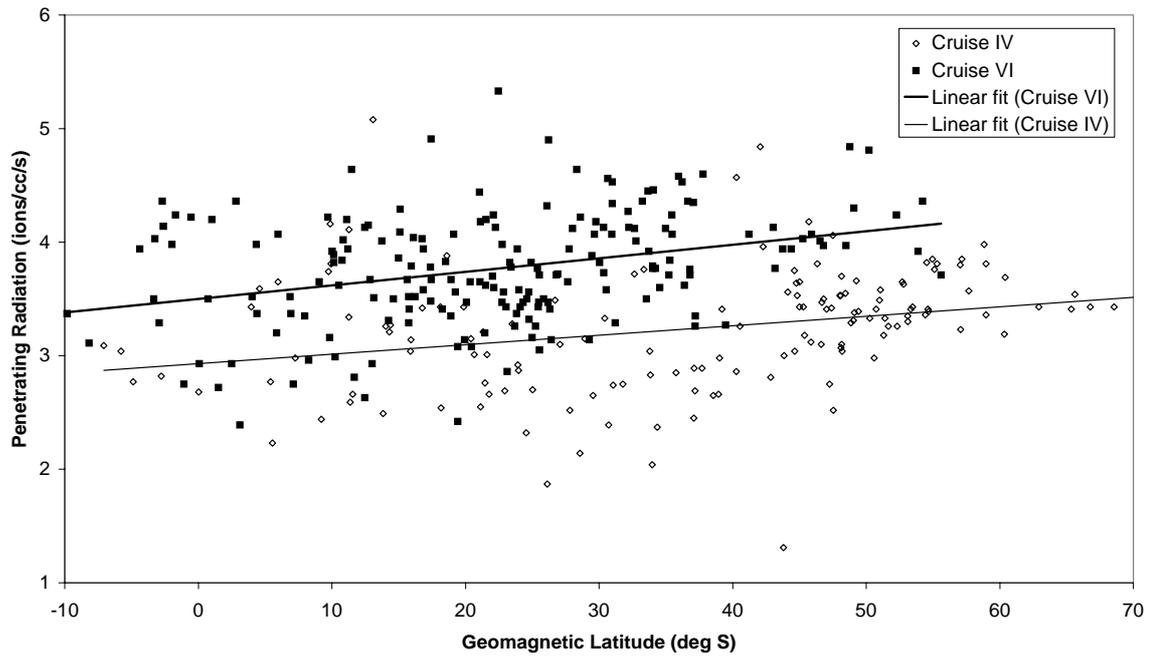

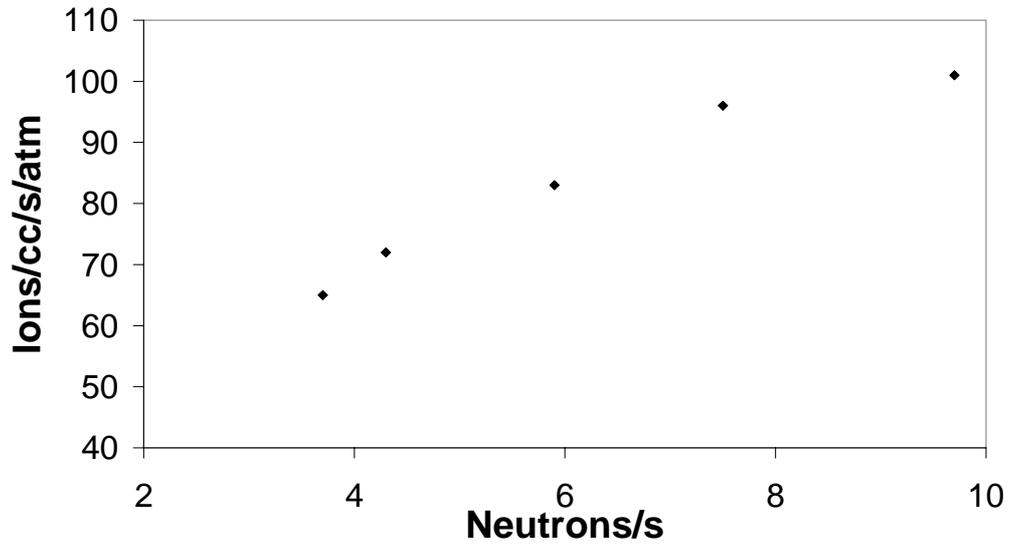

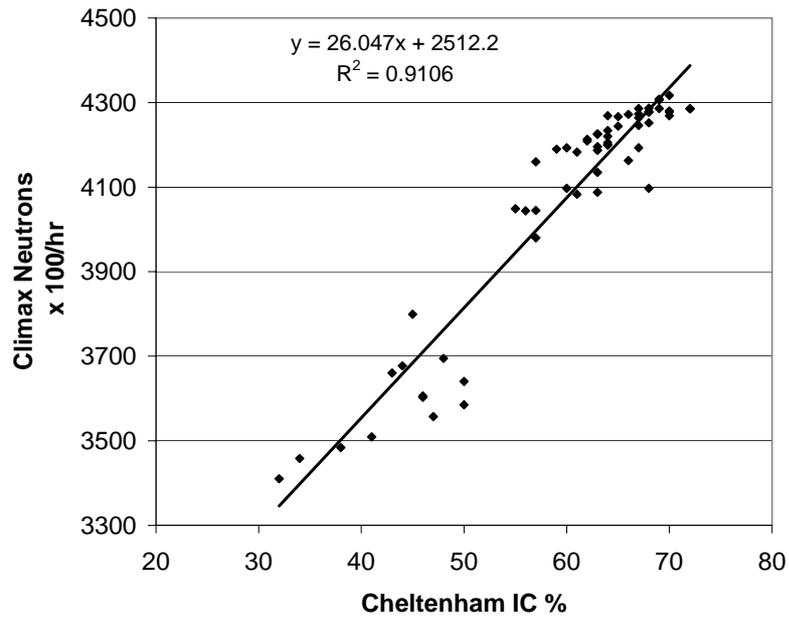

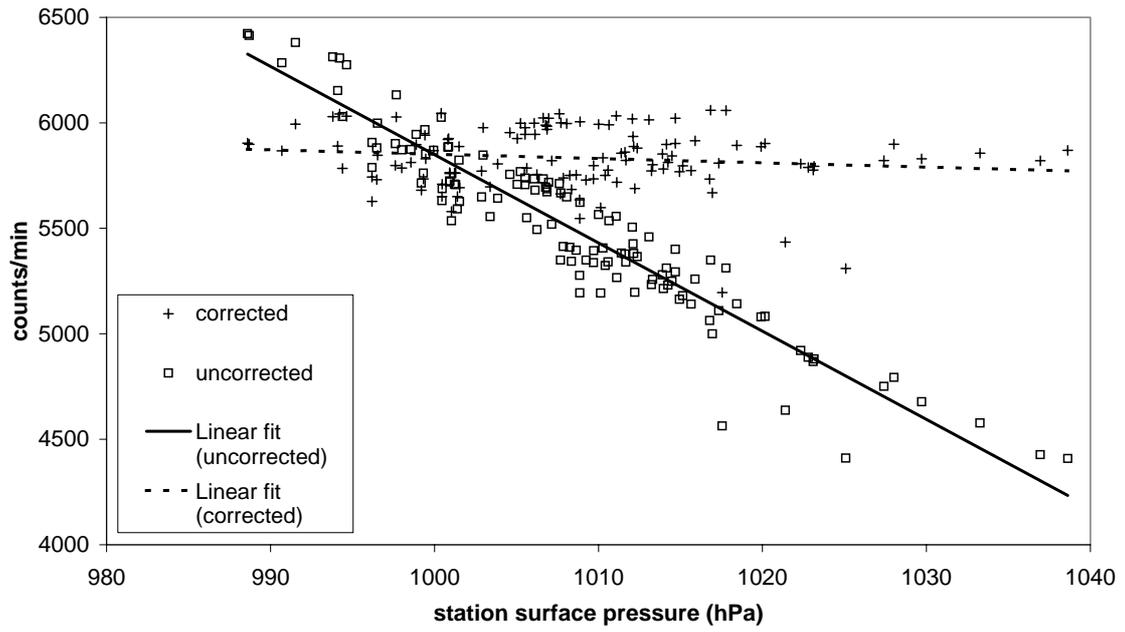

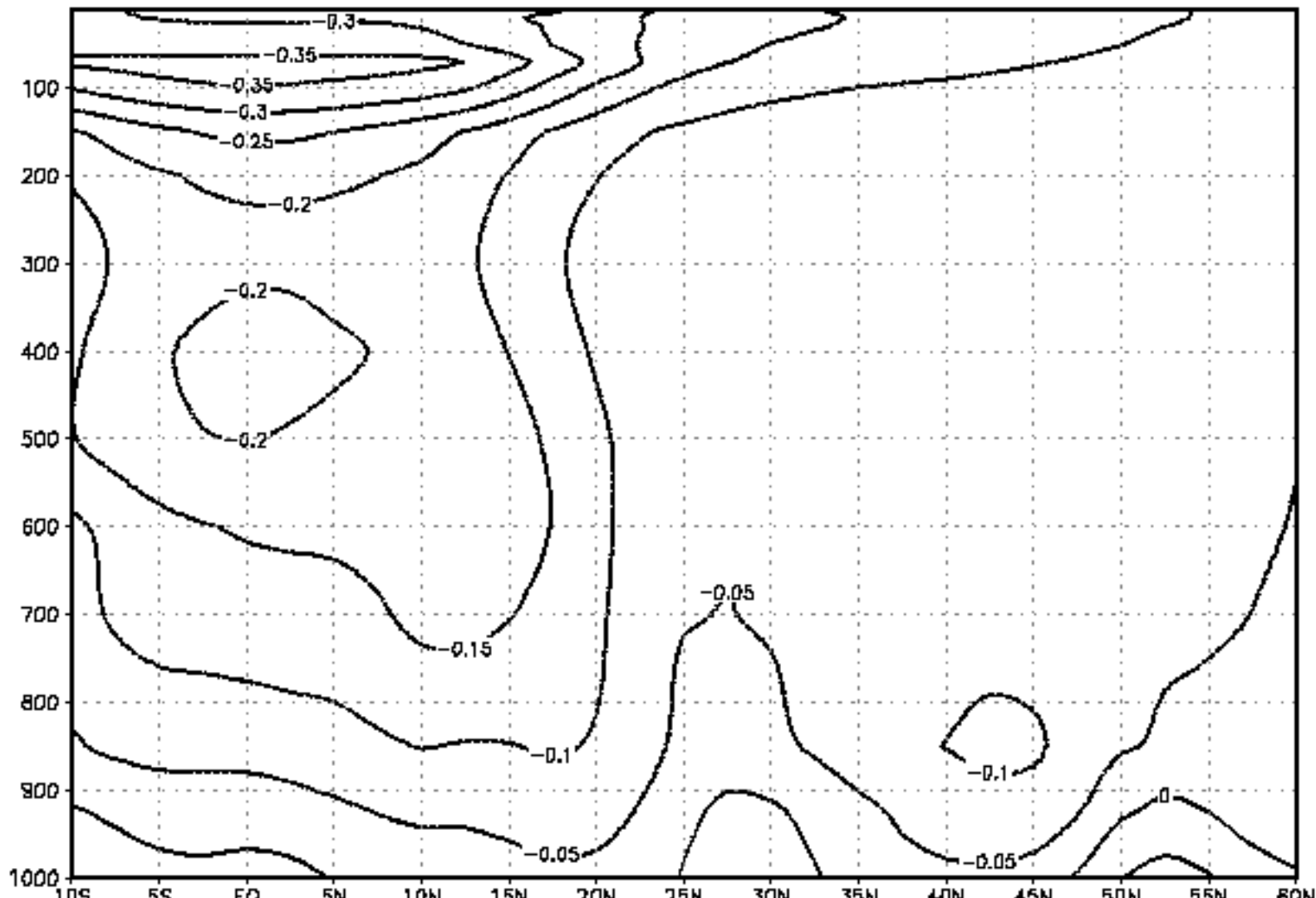

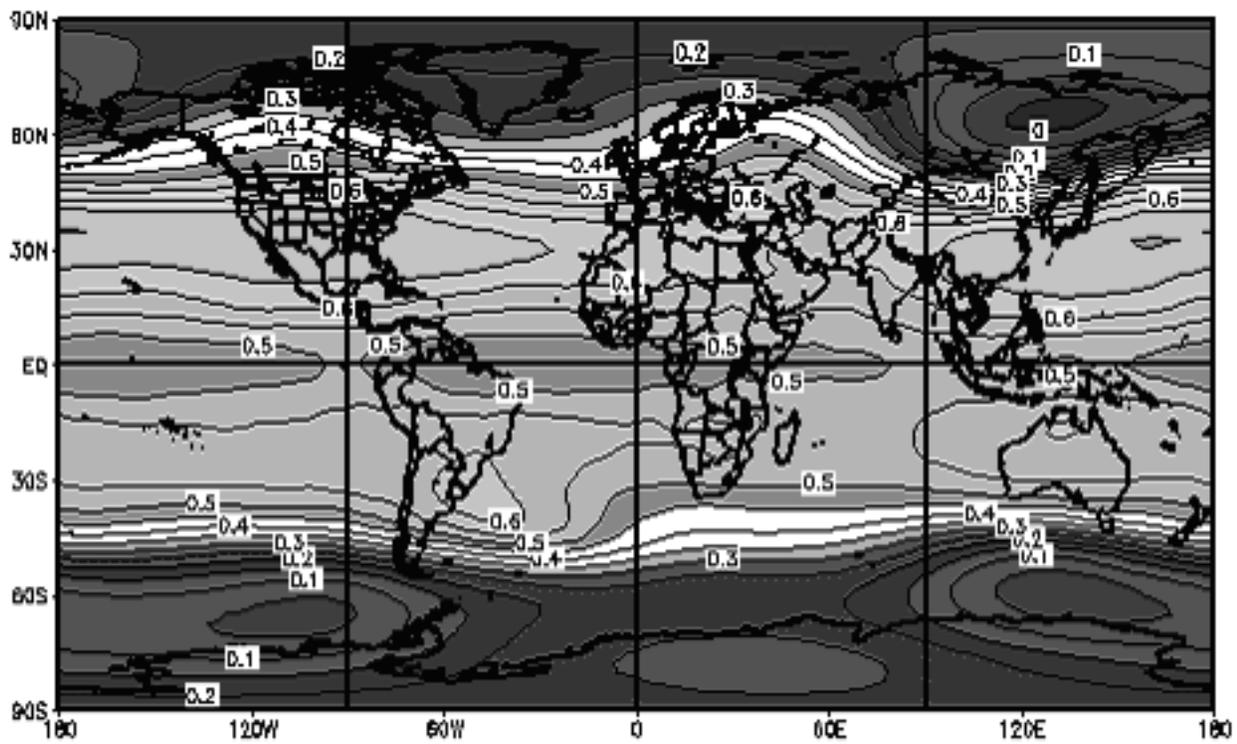
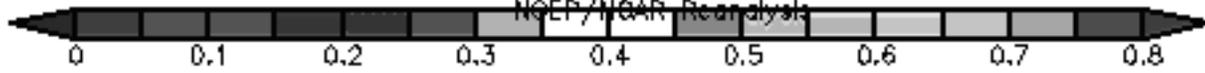

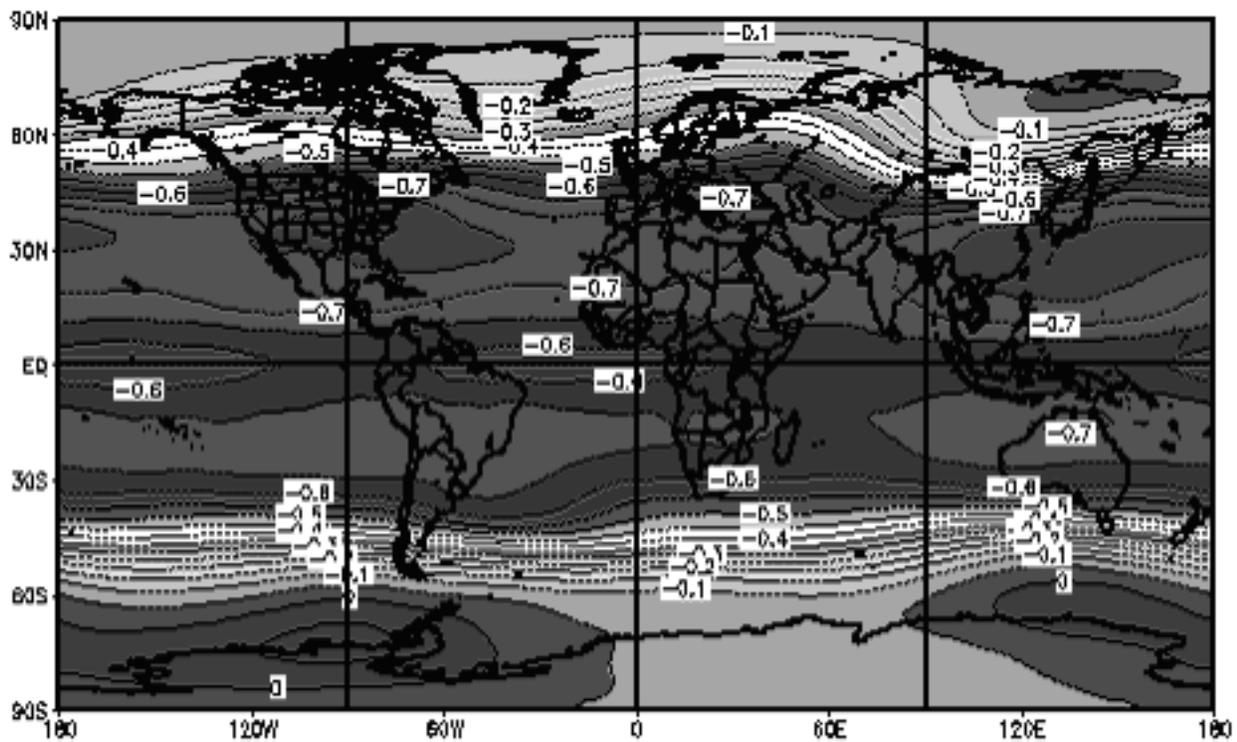